\documentclass[%
 aip,
 amsmath,amssymb,
 reprint,%
floatfix
]{revtex4-1}

\usepackage{graphicx}
\usepackage{dcolumn}
\usepackage{bm}

\usepackage[caption=false]{subfig}

\usepackage[utf8]{inputenc}
\usepackage[T1]{fontenc}
\usepackage{mathptmx}
\usepackage{etoolbox}

\usepackage{graphicx}
\usepackage[draft]{changes}
\usepackage{natbib}
\usepackage{titlesec}

\usepackage[ruled,vlined]{algorithm2e}
\usepackage[noend]{algpseudocode}
\usepackage{tabularx}
\usepackage{algpseudocode}

\usepackage{xfrac}
\usepackage{xcolor}
\usepackage{amsmath }
\definechangesauthor[name={Yin YANG},color=orange]{YY}
\definechangesauthor[name={Dominique Heitz},color=red]{DH}
\definechangesauthor[name={Ali Rahimi Khojasteh},color=blue]{AR}

\makeatletter
\def\@email#1#2{%
 \endgroup
 \patchcmd{\titleblock@produce}
  {\frontmatter@RRAPformat}
  {\frontmatter@RRAPformat{\produce@RRAP{*#1\href{mailto:#2}{#2}}}\frontmatter@RRAPformat}
  {}{}
}%
\makeatother
\begin{document}

\preprint{AIP/123-QED}

\title[Lagrangian Coherent Track Initialisation (LCTI)]{Lagrangian Coherent Track Initialisation (LCTI)}

\author{Ali Rahimi Khojasteh}
\author{Yin Yang}%
 \email{ali.rahimi-khojasteh@inrae.fr.}
 
\author{Dominique Heitz}

\affiliation{ 
INRAE, OPAALE, 17 avenue de Cucill\'e, 35044, Rennes, France 
}%

\author{Sylvain Laizet}
\affiliation{%
Turbulence Simulation Group, Department of Aeronautics, Imperial College London, Exhibition Road, London SW7 2AZ, United Kingdom
}%


\date{\today}

\begin{abstract}

Advances in time-resolved Particle Tracking Velocimetry (4D-PTV) techniques have been consistently revealed more accurate Lagrangian particle motions. A novel track initialisation technique as a complementary part of 4D-PTV, based on local temporal and spatial coherency of neighbour trajectories, is proposed. The proposed Lagrangian Coherent Track Initialisation (LCTI) applies physics-based Finite Time Lyapunov Exponent (FTLE) to build four frame coherent tracks. We locally determine the boundaries (i.e., ridges) of Lagrangian Coherent Structures (LCS) among neighbour trajectories by using FTLE to distinguish clusters of coherent motions. To evaluate the proposed technique, we created an open-access synthetic Lagrangian and Eulerian dataset of the wake downstream of a smooth cylinder at a Reynolds number equal to $3900$ obtained from 3D Direct Numerical Simulation (DNS). The dataset is available to the public. Performance of the proposed method based on three characteristic parameters, temporal scale, particle concentration (i.e., density), and noise ratio, showed robust behaviour in finding true tracks compared to the recent initialisation algorithms. Sensitivity of LCTI to the number of untracked and wrong tracks are also discussed. We address the capability of using the proposed method as a function of a 4D-PTV scheme in the Lagrangian Particle Tracking challenge for a flow with high particle densities. Finally, the LCTI behaviour was assessed in a real jet impingement experiment. LCTI was found to be a reliable tracking tool in complex flow motions, with a strength revealed for flows with high particle concentrations.

 
\end{abstract}

\maketitle



\section{\label{sec:Intro}Introduction}

\begin{figure*}
\begin{center} 

  \includegraphics[width=0.8\textwidth]{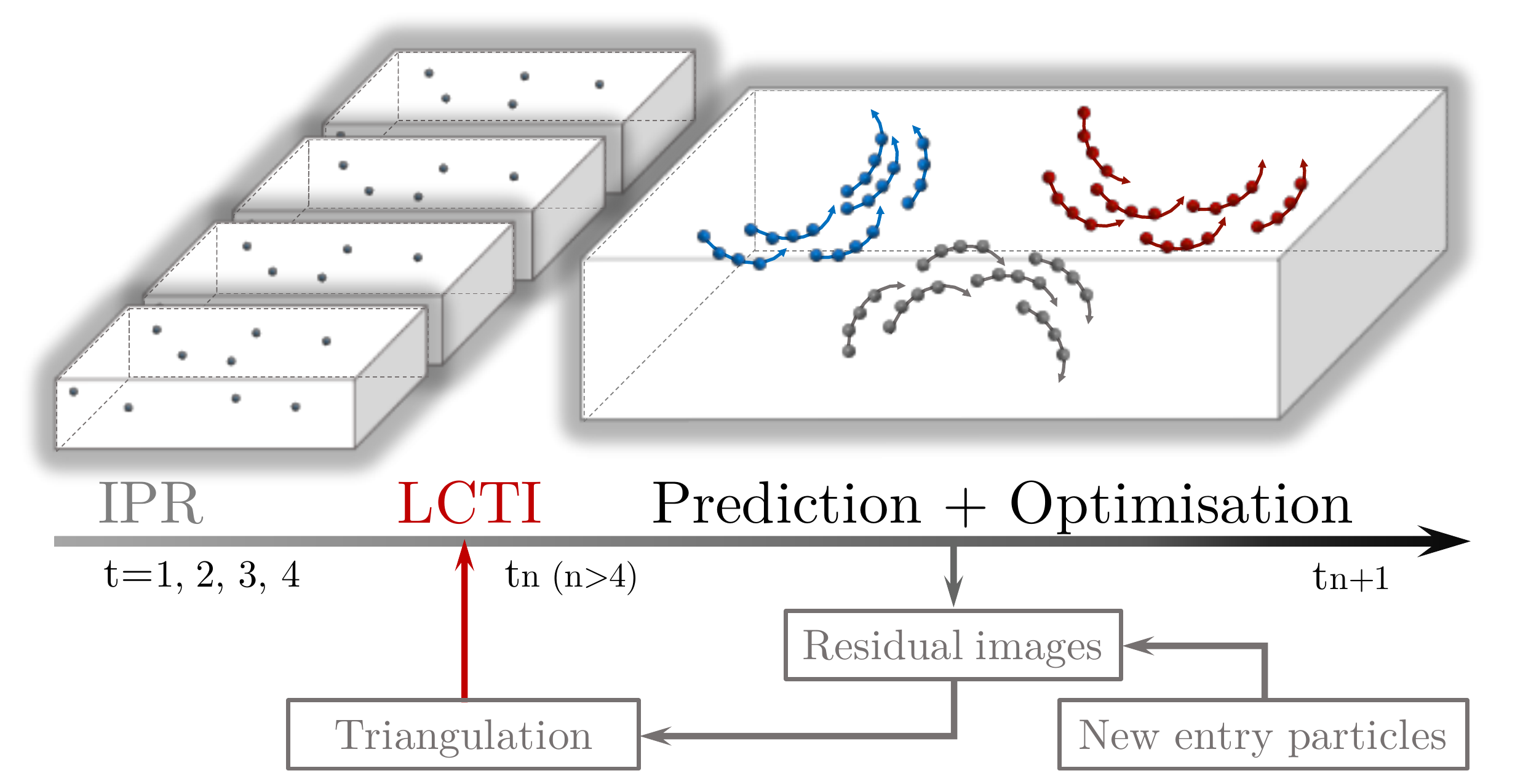}
\caption{\label{fig:fig1} 4D-PTV flowchart starting from particle reconstructions in four frames using IPR \citet{Wieneke2012}. LCTI recursively add new entry and lost tracks into the optimisation poll. }
\end{center} 
\end{figure*}

 Recent developments in the Shake-The-Box (STB) approach \citep{Schanz2016Shake-The-Box:Densities} have led to a renewed attention on time-resolved Particle Tracking Velocimetry (4D-PTV) for the study of turbulent flows. STB introduces a fast and efficient tracking idea based on a particle position prediction step followed by an image space optimisation scheme solved with "Shaking". STB was first initiated and shaped after the Iterative Particle Reconstruction (IPR) concept proposed by \citet{Wieneke2012}. A typical 4D-PTV technique comprises four recursive steps inspired by STB, as shown in Fig.~\ref{fig:fig1}.  These steps are particle reconstruction, track initialisation, prediction, and optimisation. Particle reconstruction is a process that converts with triangulation multi-view 2D particle images to particle positions in a 3D domain. However, the triangulation only works for sparse particle concentrations lower than $0.001$ particles per pixel ($\text{ppp}$). The number of ghost particles drastically increases in triangulation for higher particle concentrations due to overlapping particles. Hence, \citet{Wieneke2012} proposed IPR with an additional step to overcome the triangulation inaccuracy. In IPR, the triangulation is followed by an iterative optimisation procedure that searches for the best particle position minimising the intensity discrepancy between the original and the reprojected particle image. The reconstructed 3D positions from IPR are thereafter fed into the initialisation part (see Fig.~\ref{fig:fig1}). After the tracklets of the first few frames are built, the prediction function then estimates positions of the next time step ($t_{n+1}$) using polynomial or Wiener filter predictors \citep{Schanz2016Shake-The-Box:Densities,Tan2020IntroducingTracking}. The optimisation takes part from the predicted positions until the optimal positions at $t_{n+1}$ are found. During the prediction-optimisation phase, particles continuously enter the domain. Those new entry trajectories must be fed into the tracked poll; otherwise, there would eventually be no tracks left since all tracked particles would have left the domain for a flow with a main advection. In complex flow motions or with high particle concentrations, some particles lose their trajectories due to the optimisation failure. In this scenario, the lost particles are kept in the residual images, but their tracks will be removed from the tracked poll (see Fig.~\ref{fig:fig1}). It is vital to reconstruct those lost particles and build tracklets since an increasing number of lost tracks will lead to the divergence of the tracking algorithm. To this end, the track initialisation attempts to build tracklets for three types of particles, particles in the first four/five frames, new entries, and lost particles (see Fig.~\ref{fig:fig1}). 

4D-PTV methods like STB and Kernelized Lagrangian Particle Tracking (KLPT, \citet{yang:hal-03212696}) require an appropriate and reliable number of initialised true tracks at every time step. Otherwise, the tracking process fails to reconstruct trajectories for the majority of particles. Such a failure illustrates the importance of implementing a robust multi-frame track initialisation technique to prevent 4D-PTV divergence, particularly in dense and complex situations. The idea of initialising a possible track in four frames, known as four frame best estimate (4BE), has been widely used in LPT/PTV studies \citep{Malik1993ParticleTracking,Ouellette2006AAlgorithms,Dou2018Particle-pairVelocimetry,Clark2019ATracking,Machicoane2019RecentResearch}. Four frame tracking methods with simple nearest neighbour particle matching have been introduced for low density, and smooth flow behaviours \citep{Malik1993ParticleTracking}. New studies recently improved four frame track initialisation performance, including four-frame best estimate \citep[4BE-NNI method]{Ouellette2006AAlgorithms} by looking for the nearest neighbours in sequential frames until a unique solution is found or Enhanced Track Initialisation \citep[4BE-ETI method]{Clark2019ATracking} by looking for all track possibilities with an adjustable search volume \citep[see also][]{Machicoane2019RecentResearch}. \citet{Dou2018Particle-pairVelocimetry} proposed initialising with two nearest candidates in a similar spirit, then kept predicting and particle matching in the next two following time steps (four frames in total) until a unique match is found. \citet{Cierpka2013Higher2012} have shown that the four frame methodology could be extended to multi-frame tracking with the combination of neighbour possibility and temporal prediction in sequential steps. The most straightforward prediction function is the linear predictor that can be calculated from the position difference of every two possible matches. The use of a linear predictor improved the probability of finding true tracks as well as reducing the computation time by having a targeted search volume. Some studies also suggest applying the prior Particle Image Velocimetry (PIV) velocity field as a predictor  \citep{Cardwell2011AFlows, Schanz2016Shake-The-Box:Densities}. Although this idea is applicable to 2D and 3D studies, extracting a 3D PIV velocity field is expensive due to its spatial resolution, uncertainty, and complex experimental issues.

For complex flow motions and experiments with high particle densities, a further step for the track validity check is required to avoid false detection. \citet{Guezennec1994AlgorithmsVelocimetry} originated this validity check as a self-coherency algorithm with the concept of path coherence. Their technique minimises a penalty function from all possibilities in five frames. In their study, a possible track is a coherent path if it is a smooth trajectory in position, velocity, acceleration, and rate of acceleration. This spatial and temporal self-consistency penalty function only focuses on a single track behaviour. Recent four frame based techniques also performed similar self-consistency approaches. As an example, \citet{Dou2018Particle-pairVelocimetry} checked if the velocity differences between two frames exceed a certain threshold to validate a possible track. With a comparative approach, \citet{Ouellette2006AAlgorithms} controlled the acceleration change instead of velocity. These post-treatments rely on spatial and temporal filtering of the trajectories to avoid false tracks. However, as complex flows often feature high gradient motions, over-smoothing dynamics could lead to a quality degeneration of the reconstructed field. The degeneration problem is that there is no direct learning based on the physics of flow included in the classical four frame schemes, which brings more challenges for complex and high-density flow motions.

We argue that when a track can not be reconstructed successfully solely due to ambiguities caused by overlapping and multi possibilities, it is always beneficial to extract more information from its neighbourhood. If the reconstructed tracks are available in the neighbourhood, we can use this information to gain a better insight into the target particle's potential behaviour and eventually solve the ambiguity problematic associated with the initialisation. As we are dealing with the fluid flow, assuming a constant solid local rigid neighbourhood is naive and erroneous, it is essential to consider a coherent neighbourhood where both the target particle and neighbour particles share the coherent motion. 

In this study, we propose a novel Lagrangian Coherent Track Initialisation (LCTI) technique for finding tracks in four (or multi) frames that belong to clusters of coherent motions. We apply Lagrangian Coherent Structures \citep[LCS]{Haller2015LagrangianStructures} to distinguish coherent and non-coherent neighbour trajectories. The LCS, also known as the skeleton of flow, determines separatrix lines or surfaces that divide flow structures into different coherent regions. We use Finite Time Lyapunov Exponent (FTLE) that is the most common method in quantifying these separatrices boundaries \citep{Balasuriya2020HyperbolicStretching}.
The paper is organised as follows: we first present the LCTI algorithm based on Lagrangian coherency in four frames in section \ref{sec:2}. We then address in section \ref{sec:3} the concept of clusters of coherent particles, and how to quantify the coherent trajectories. In the following section \ref{sec:4}, we discuss how to create synthetic datasets in the wake behind a smooth cylinder at a Reynolds number equal to $3900$ (based on the cylinder diameter and freestream velocity) because it is an effective tool to evaluate particle tracking techniques \citep{Lecordier2004TheS.I.G.} and has been utilised widely in recent PTV studies \citep{Schanz2016Shake-The-Box:Densities,Tan2020IntroducingTracking,Patel2018RapidFields,Clark2019ATracking}. LCTI robustness, accuracy and sensitivity analyses are also illustrated in terms of particle concentration, temporal scale, and noise level. Finally, we demonstrate the LCTI performance for the LPT challenge wall-bounded wake flow behind a cylinder \citep{Cfdforpiv.dlr.de20203rdTracking} and a jet impingement experiment in sections \ref{sec:4} and \ref{sec:5}. 

\section{Four Frame Coherent Track Initialisation}
\label{sec:2}

The current initialisation technique tries to find coherent tracklets in four frames. It should be noted that particles of a cluster are coherent if they spatially behave together over a finite time. A starting step is required in LCTI for the first time step $t_1$, where there is no neighbour track information. It can be done by a classic four-frame scheme with a narrow threshold to index the most reliable tracks. It is assumed that a track is relatively reliable if it has comparable small velocity and acceleration standard deviations in four time steps to avoid false tracks. The standard deviation of the particle image intensity can also determine whether a possible track is reliable. In practice, the LCTI steps can be listed as the following algorithm,

\begin{algorithm}[H]
\label{alg:lcti}
\caption{LCTI}
\begin{algorithmic}[1]
\State Index possible candidates inside the search volume at $t_2$;
\State Two consecutive predictions and candidate matching for $t_3$ and $t_4$;
\State Index possible tracks from $t_1$ to $t_4$;
\State Coherency check with neighbour tracks for each possible track;
\State Index the most coherent track.
\end{algorithmic}

\end{algorithm}


\begin{figure*}
\begin{center} 
  \includegraphics[width=0.8\textwidth]{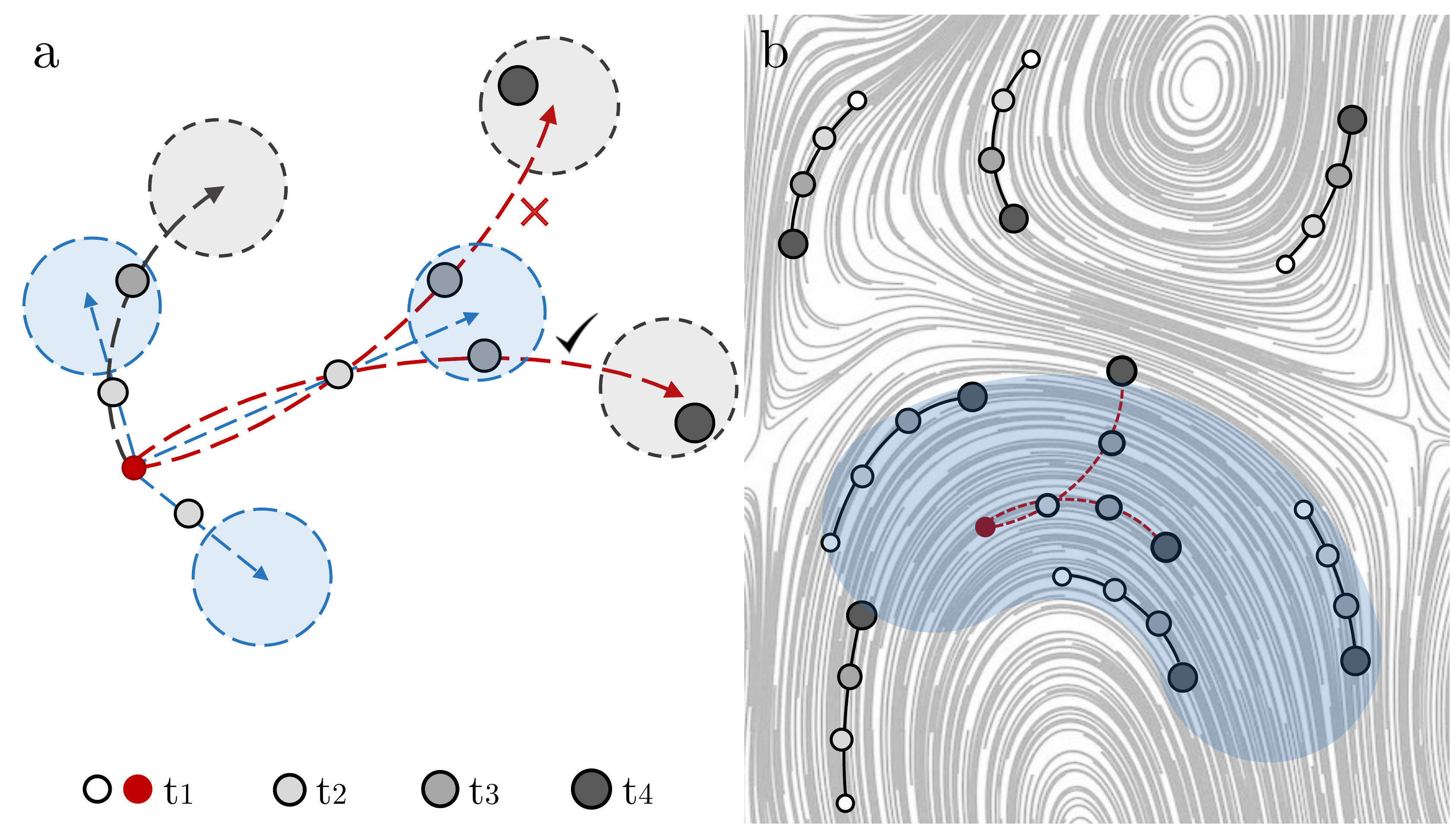}
\caption{\label{fig:fig2} Schematic view of the LCTI algorithm when two possible four frame solutions exist. a) LCTI four frame algorithm considering all possible neighbour candidates at $t_2$ followed by linear predictions (blue dash line arrows). Candidate matching at time step $t_3$ inside first search volume (blue circle). Second order prediction (red and black dash line arrows) to match possible candidates at $t_4$ inside second search volume (grey circle). b) Coherency check between two possible track matches and neighbour coherent motion }
\end{center}
\end{figure*}

Referring to the LCTI algorithm~\ref{alg:lcti}, we need to define the search volumes to index possible candidates at each time step. \citet{Clark2019ATracking} enhanced the probability of finding true tracks by applying adjustable anisotropic search volumes as a function of mean flow direction. Anisotropic means that if the mean flow (obtained from the predicted velocity) is dominant in one direction, the search volume in that direction is larger than in the other directions. Adjustable search volumes introduce local spatial motions (i.e., physics-based information) into four frame schemes, which can significantly tackle the high gradient threshold issues. On the other hand, using the adjustable search volume limits the number of possible candidates, avoiding non-coherent solutions by following the local spatial motion. The search volume in LCTI is based on the local maximum displacement map calculated from neighbour particles. Therefore, the first search volume is computed as a function of neighbour maximum displacements in each spatial direction between $t_1$ and $t_2$ as shown in Fig.~\ref{fig:fig2}a. Then, every neighbour particle inside the search volume at $t_2$ is a candidate. These candidates are in one of the following categories: the true position of the target particle at $t_2$, the true position of other undetected tracks, and noise (i.e., false particle). Afterwards, a linear predictor (blue dash line arrows in Fig.~\ref{fig:fig2}a) between the target particle at $t_1$ and the possible candidate at $t_2$ is performed for every possible match. Similarly, the second search volume around the predicted position determines which particles are more likely to be in the true position at $t_3$. A possible track is removed if there is no candidate inside the search volume. The process is repeated for the next time step with a higher order prediction function (red and black dash line arrows in Fig.~\ref{fig:fig2}a). A unique four frame solution is expected for flows with low velocity and acceleration gradients or low particle concentrations. When more than one solution exists, LCTI selects the most coherent track to solve the ambiguities, as shown in Fig.~\ref{fig:fig2}b. Otherwise, a particle can spatially meet a group of other particles with no coherency link between them. We recall here that coherent refers to a group of particles that are having the same Lagrangian behaviour spatially and temporally. A function is therefore required to determine coherent and non-coherent clusters of particles locally. More details and principles on the coherent motion of particles are discussed in Section \ref{sec:3}.

\section{Coherent Track Detection}
\label{sec:3}

\begin{figure}
\begin{center} 

  \includegraphics[width=0.5\textwidth]{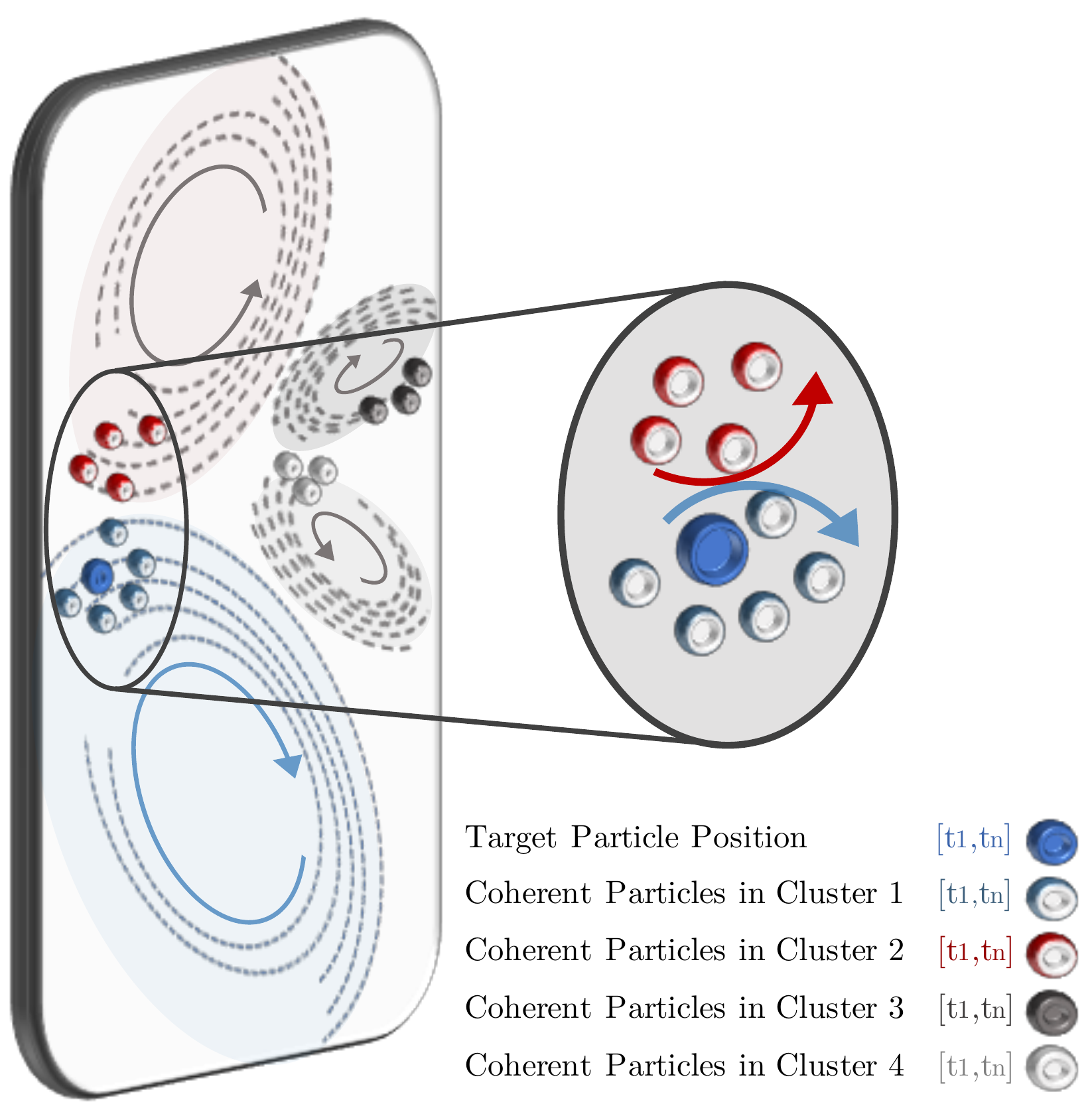}
\caption{\label{fig:fig3} 2D schematic of particle motions inside vortices. Each colour belongs to a group of coherent clusters. The target dark blue particle with coherent neighbour particles is located in a clockwise vortex (blue cluster) while non-coherent particles belong to different clusters. The target particle is non-coherent with neighbour particles in the red cluster.}
\end{center} 
\end{figure}

Recently, Lagrangian Coherent Structures (LCS) have been applied in PIV/PTV experiments for flow structure analyses (see, e.g., \citet{Krishna2018FlowfieldWing}). However, previous studies have not yet combined the LCS extraction with the velocimetry algorithms and mostly focused on using LCS as a post-processing tool, to the best of our knowledge. Several methods have been proposed to identify LCS by looking for separatrix regions in time \citep{Haller2015LagrangianStructures,Hadjighasem2016Spectral-clusteringDetection}. Separatrices exist in boundaries (i.e., ridges) between different structures. A schematic view of the boundaries between vortices in a 2D isotropic homogeneous turbulent flow is shown in Fig.~\ref{fig:fig3}. Multi clusters of particles spatially exist in the vicinity of the target particle (the dark blue particle in Fig.~\ref{fig:fig3}). All red and blue particles are neighbours of the target particle. However, the trajectories of each coloured cluster temporally evolve in separated directions. LCS can be used to determine if a spatially neighbour particle is coherent or non-coherent over a temporal scale.


Suppose the flow is dominated by coherent structures such as in a 2D isotropic homogeneous turbulence illustrated in Fig.~\ref{fig:fig3}, the global LCS analysis can extract meaningful boundaries between structures. Difficulties in interpretation arise, however, when the flow carries 3D complex motions and numerous local structures. We, therefore, suggest local coherent structure extractions instead of a global calculation for which only coherent clusters and boundaries over neighbour trajectories are computed. Therefore, the complexity of the global LCS view is simplified into a small number of clusters around the target particle, such as in Fig.~\ref{fig:fig3}. In the local view, curve or surface boundaries divide the local spatial area into discrete regions with different dynamic motions \citep{Samelson2012LagrangianStrain}, and motions across these boundaries are negligible \citep{Shadden2005DefinitionFlows}. Furthermore, the LCS boundaries can move, evolve, and vanish in spatial space as the flow pattern changes temporally. 


In the local Lagrangian frame, separatrices can be obtained from Finite Time Lyapunov Exponent (FTLE) by measuring the amount of stretching between the target particle and its neighbour particles over finite time \citep{T.Alligood1996CHAOS:Systems,Raben2014ComputationData}. \citet{Raben2014ComputationData} showed that the normalised average error and normalised root-mean-squared (RMS) error of the FTLE map decreases with increased particle concentration. This trend is favourable because ongoing PIV/PTV experiments consistently succeed in achieving higher particle concentrations. Meanwhile, it is less likely to have ambiguities due to multi possible solutions in low particle concentration cases. As a result, there is no critical need for the coherency check in low particle concentration cases. 

As discussed in section \ref{sec:2}, if a possible track is coherent with its neighbour tracks, it will be indexed into the tracked poll. To compute FTLE, we assumed that the local flow map is the particle trajectory in finite time from $t_0$ to $t_0+T$. The idea is to analyse the spatial displacements between the target particle and its neighbours. The flow map of a single particle can be formulated as
\begin{equation}\label{flowmap}
\emptyset_{t_0}^{t_{0+T}}\left(x\right)\-:\ x(t_0)\ \rightarrow\ x(t_0+T)\ ,
\end{equation}
where $x(t_0)$ is the starting position of the interval time $T$, and $x(t_0+T)$ is the final position. Difference between the flow maps of the target particle $\emptyset_{t_0}^t\left(x_0^p\right)$ and its neighbour $\emptyset_{t_0}^t\left(x_0^n\right)$ would result in a vector displacement as following,
\begin{equation}\label{vector displacement}
    {\delta x(t)=\emptyset}_{t_0}^t\left(x_0^p\right)-\emptyset_{t_0}^t\left(x_0^n\right).
\end{equation}

This vector displacement contains transformation between the initial and final positions of two particles. Equation \ref{vector displacement} can be linearised by using the first term of the Taylor series of $\emptyset_{t_0}^t\left(x_0^n\right)$ expanded about $x_0^p$:
\begin{equation}\label{deformation}
\delta x(t) \approx \frac{\partial\emptyset_{t_0}^t\left(x_0^p\right)}{\partial x_0}\delta x_0 ,
\end{equation}
where $\delta x_0=x_0^p-x_0^n$. The state-transition matrix $\partial\emptyset_{t_0}^t\left(x_0^p\right)/\partial x_0$ is also known as deformation gradient tensor $\nabla\emptyset_{t_0}^t\left(x_0\right)$ \citep{Shadden2005DefinitionFlows}. The deformation tensor carries valuable information including the rate of expansion, compression and rotation. The magnitude of $\delta x(t)$ is
\begin{align}
\label{FTLE_derivative}
|\delta x(t)| &=\sqrt{\delta x(t)\delta x(t)}\nonumber \\ &=\sqrt{(\nabla\emptyset_{t_0}^t(x_0)\delta x_0)(\nabla\emptyset_{t_0}^t(x_0)\delta x_0)} \nonumber \\
&=\sqrt{\delta x_0(\nabla\emptyset_{t_0}^t(x_0)\nabla\emptyset_{t_0}^t(x_0))\delta x_0},
\end{align}
where we define
\begin{equation}\label{Delta}
    \Delta=\nabla\emptyset_{t_0}^t(x_0)\nabla\emptyset_{t_0}^t(x_0).
\end{equation}

$\Delta$ is a symmetric positive definite matrix, also known as the right Cauchy-Green deformation tensor \citep{Shadden2005DefinitionFlows} with three real and positive eigenvalues in a 3D domain over finite time.


As mentioned before, FTLE measures the rate of stretching between the target particle and its neighbour. The maximum eigenvalue of the Cauchy-Green tensor $\lambda_{max}(\Delta)$ shows the expansion or separation in finite time. Furthermore, the eigenvector corresponding to $\lambda_{max}(\Delta)$ represents the direction of the separation. Eventually, the magnitude of the maximum displacement can be written as
\begin{equation}\label{maximum magnitute}
    |\delta x(t)_{max}|=\sqrt{\delta x_0\lambda_{max}\left(\Delta\right)\delta x_0}=\sqrt{\lambda_{max}\left(\Delta\right)}|\delta x_0|,
\end{equation}
and the FTLE value $\Lambda_{t_0}^t$ is defined as 
\begin{equation}\label{FTLE}
\Lambda_{t_0}^t=\frac{1}{\left|T\right|}\sqrt{\lambda_{max}\left(\Delta\right)}=\frac{1}{\left|T\right|}log\left (\frac{\delta x(t)}{\delta x(t_0)}\right ).
\end{equation}

A lower FTLE value means the neighbouring particle is acting similarly, with no sign of separation with the target particle over the finite time. High values in the FTLE field show the existence of the LCS ridges that divide the local area into different clusters of coherent particles. With this formulation, it is possible to index a group of neighbour particles as coherent or non-coherent with the target particle. As we discussed in section \ref{sec:2}, the LCTI algorithm checks if the Lagrangian coherency is valid for each possible four frame tracks to avoid non-coherent reconstructions. Assuming two possible matches exist for the target particle (see Fig.~\ref{fig:fig2}b), we start by fitting a smooth curve over each known neighbour tracks to reduce the noisy reconstruction effect on the coherency detection. Then LCTI locally computes the FTLE map over the fitted tracks without considering two possible matches. If the FTLE map shows local separations, neighbour tracks in the same cluster with the target particle are classified as coherent neighbours. As illustrated in Fig.~\ref{fig:fig3}, the local region is divided into two blue and red clusters. Only the neighbouring tracks inside the blue cluster are coherent with the target particle. All neighbour tracks are coherent neighbours if no separation is detected. To quantify a threshold for the FTLE ridge detection, we assessed the FTLE map for the case of 2D homogeneous isotropic turbulence given by Direct Numerical Simulation (DNS). We found that values above the threshold $0.25$ are optimal criteria to estimate the FTLE ridge positions. This threshold was in agreement with studies using global FTLE ridge calculation in the range of $50 \% - 80 \%$ of the maximum FTLE value \citep{Lipinski2010AStructures}. There are valuable studies in ridge detection algorithms with extensive computation costs that can be employed instead of using a constant FTLE threshold (see, e.g., \citet{Shadden2005DefinitionFlows}). After the coherent neighbour determination, LCTI checks the FTLE value for each possible match and neighbours. Finally, the most coherent match with the coherent neighbours will be indexed. This process continues iteratively until no track is found to be coherent with the tracked poll. 

\section{Synthetic evaluation}
\label{sec:4}

To quantify and distinguish the LCTI performance from other schemes, we first applied the tracking process to synthetic data. In this section, we consider a set of synthetic data for the wake flow downstream of a smooth cylinder. We first describe the creation of the Lagrangian data by transporting the synthetic particles in the Eulerian velocity volume. Thereafter, we discuss how to set characteristic parameters for the synthetic data, including 1- Particle concentration, 2- Temporal scale, and 3- Noise level. We also address accuracy analyses of LCTI in comparison with other schemes, followed by the sensitivity analysis of LCTI to every characteristic parameter. Finally, we discuss the LCTI performance in the LPT challenge for the wall-bounded wake flow behind the cylinder \citep{Cfdforpiv.dlr.de20203rdTracking}.   

\subsection{Synthetic data creation}
\label{sec:4_1}
\subsubsection{Particle transport}
 
An open-source Direct Numerical Simulation (DNS) code, named \texttt{Incompact3d}, was employed to simulate the wake downstream of a smooth cylinder at a Reynolds number equal to $3900$. \texttt{Incompact3d} is part of the \texttt{Xcompact3d} framework dedicated to the study of turbulent flows using supercomputers \citep{Bartholomew2020Xcompact3D:Mesh} and it is dedicated to the incompressible Navier-Stokes equations. It is based on sixth-order finite-difference schemes on a Cartesian mesh for spatial discretisation and, for the present simulation, a third-order Adams–Bashforth scheme for the time advancement. The main originality of \texttt{Incompact3d} is that
the Poisson equation for the incompressibility of the velocity field is fully solved in spectral space via the use of relevant 3D Fast Fourier transforms (FFTs). With the help of the concept of modified wavenumber, the divergence free condition is ensured up to machine accuracy.  The pressure mesh is staggered from
the velocity one by half a mesh to avoid spurious pressure oscillations observed in a fully collocated approach. The simplicity of the mesh allows an easy
implementation of a 2D domain decomposition based on pencils. More information of the code can be found in  \citet{Laizet2009High-orderAccuracy,Laizet2011Incompact3d:Cores}. In the present work, the smooth cylinder is modelled using a customised immersed boundary method with an artificial flow inside the cylinder to ensure the smoothness of the velocity field while imposing a no-slip boundary condition at the cylinder. Comparisons with experiments for this test case can be found in \citet{Chandramouli2018CoarseUncertainty}. A DNS performed with high-order schemes is the natural choice for the present work, with the highest possible spatial and temporal accuracy for Eulerian/Lagrangian data. Velocity and pressure Eulerian data are stored as well as Lagrangian trajectories for a duration of three vortex sheddings. These data will be made in open access and available to the public.

Lagrangian particle transport is a process to extract trajectories from an Eulerian velocity volume \citep{vanSebille2018LagrangianPractices}. In this study, the Eulerian velocity volume is a fraction of flow around the smooth cylinder of diameter $D$, starting from $1D$ upstream of the cylinder centre to $3D$ downstream in the streamwise direction to reduce the computation costs and only focus on the most complex flow regions. Every DNS snapshot (3 velocity components and pressure field) requires roughly $6 \ \text{Gb}$ of storage. Eulerian data were only recorded every $10$ DNS time steps due to storage limitation. To quantify the accuracy of particle transport, we compared trajectories obtained from every $10$ with every $1$ DNS time steps in a reduced domain. It was found that the mean of position error after $1000$ DNS time steps are $3.28$ times larger than the Kolmogorov length scale $\eta$. The standard deviation of position error was $\sigma_{STD} = 0.017 \ \eta$. It means that the trajectories we built are not exactly ground truth due to small position deviation compared with every $1$ DNS time step. However, it has no impact on the following assessment of the method we propose. More details of the current synthetic data can be found in \citet{Khojasteh2021Lagrangian3900}. The domain of interest $L_x \times L_y \times L_z$ has a dimension of $4D\times\ 2D \times\ 2D$ (from 1D upstream the cylinder). The original computational domain $20D\times\ 20D \times\ 6D$ is discretised with $n_x \times n_y \times n_z = 1537 \times 1025 \times 256$ mesh nodes with a time step of $0.00075 D/U_\infty$ (where $U_\infty$ is the freestream velocity). The mesh is uniform in the streamwise and spanwise directions while a non-uniform mesh is used in the vertical direction, with a mesh refinement towards the centre of the cylinder. The finest mesh size in the vertical direction is $\Delta y_{min}=0.00563D$.
Data are saved for $1500$ snapshots every $10$ DNS time steps $dt_{\rm DNS}$ corresponding to a duration of nearly three vortex sheddings. From these data, particles are transported according to the following steps:
First, synthetic random particles are created in the domain of interest for the first time step; Second, accurate time stepping and spatial interpolation schemes are used to transport scattered random particles in time and space \citep{vanSebille2018LagrangianPractices}. We estimated the Lagrangian particle velocities by spatial trilinear interpolation from $8$ nearest Eulerian velocities at each time step. The synthetic particle positions are updated in time using the 4th order Runge-Kutta scheme. Finally, we collected the ground truth Lagrangian trajectories by repeating the transport process for each time step. 



\begin{figure}
\begin{center} 
  \includegraphics[width=0.5\textwidth]{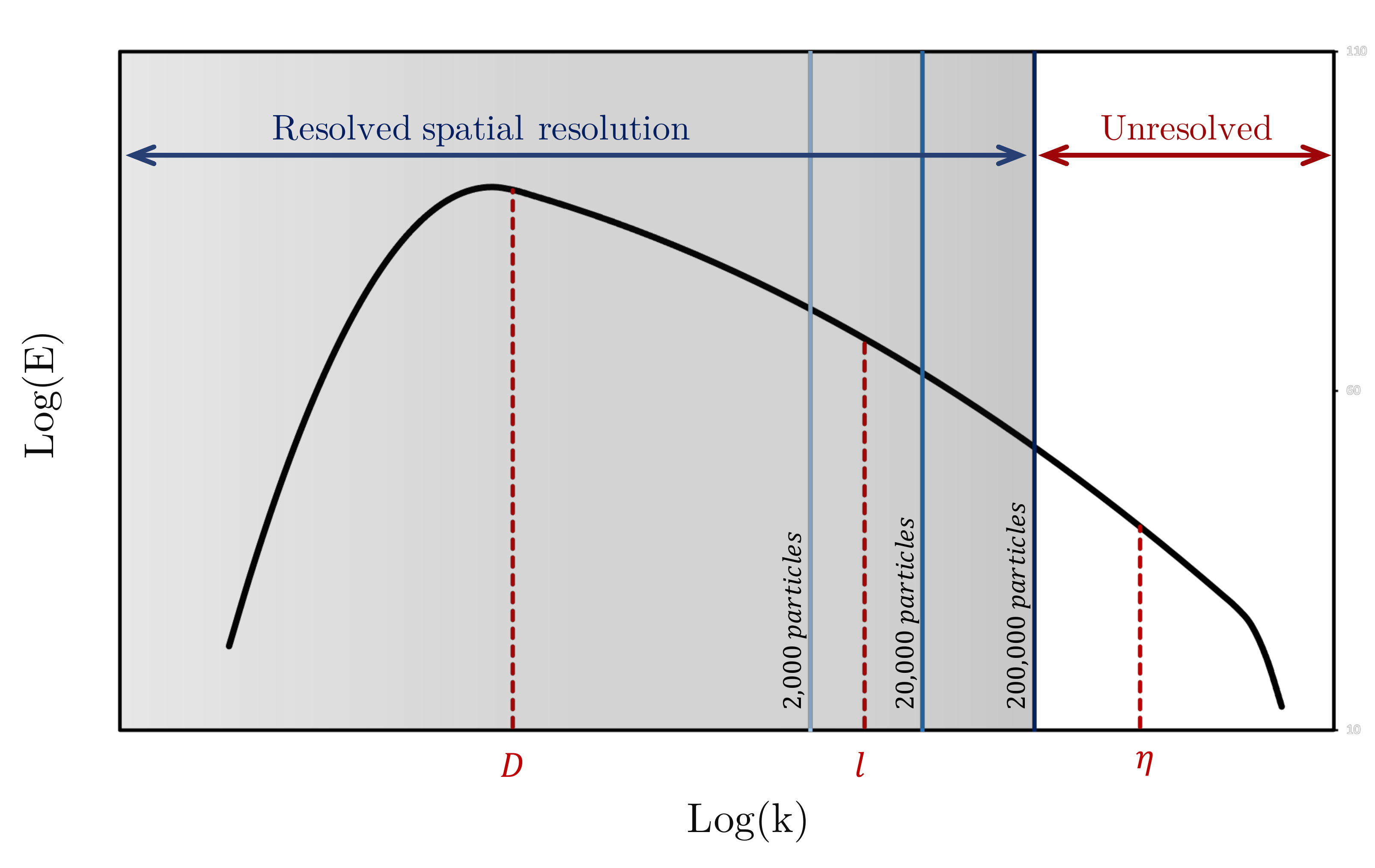}
\caption{\label{fig:fig4} Schematic view of energy spectrum in turbulence. Resolved and unresolved spatial resolutions of $2,000$, $20,000$, $200,000$ particles in blue lines with respect to integral($D$), Taylor($l$) and Kolmogorov($\eta$) length scales in red dash-lines }
\end{center}
\end{figure}



 





 




\subsubsection{Particle concentration}

In most PIV/PTV studies, particle concentration refers to the number of particles per pixel ($\text{ppp}$) that is principally an image-related parameter. The relation between the particle concentration and the turbulence length scales determines the maximum achievable spatial resolution. Therefore, it is crucial to address the number of particles corresponding to Kolmogorov ($\eta$), Taylor ($l$), and Integral ($D$) length scales. To quantify these relations, we used two volumetric metrics, namely, particles per cubic Kolmogorov scale ($\text{pp}\eta^3$) and particles per cubic Integral scale ($\text{pp}D^3$), instead of $\text{ppp}$. We chose three low, moderate, and high particle concentrations, respectively equal to $2,000$, $20,000$, and $200,000$ particles (see Table \ref{tab:table1}). Fig.~\ref{fig:fig4} shows the turbulence energy spectrum marked by the three particle concentrations (blue lines) compared with the turbulence length scales (red dash lines). Referring to the DNS data in Section  \ref{sec:4_1}, we estimated the Kolmogorov length scale as $D/\eta\sim Re^{3/4}$ that is almost $2.8$ times smaller than the average mesh size in the $y$-direction. Since the cylinder is inside our domain, it is necessary to subtract the volume of the cylinder before computing $\text{pp}\eta^3$. Accordingly, the spatial resolution for the case with $2000$ particles leads to unresolved Taylor and Kolmogorov scales. However, there are enough particles to resolve the large flow motions. Taylor length scales can be resolved by adding an order of magnitude more particles, but the smallest scales are still unresolved with $10^{-4} \ \text{pp}\eta^3$. As shown in Fig.~\ref{fig:fig4}, in our case, even $200,000$ particles with $10^{-3} \ \text{pp}\eta^3$ are not enough to resolve the smallest turbulence length scale at Reynolds number equals to $3900$. It is possible to reach up to the DNS spatial resolution numerically, but very few PTV experimental studies have reached more than $200,000$ particles in practice.

\subsubsection{Temporal scale}

\begin{figure}
\begin{center} 
  \includegraphics[width=0.5\textwidth]{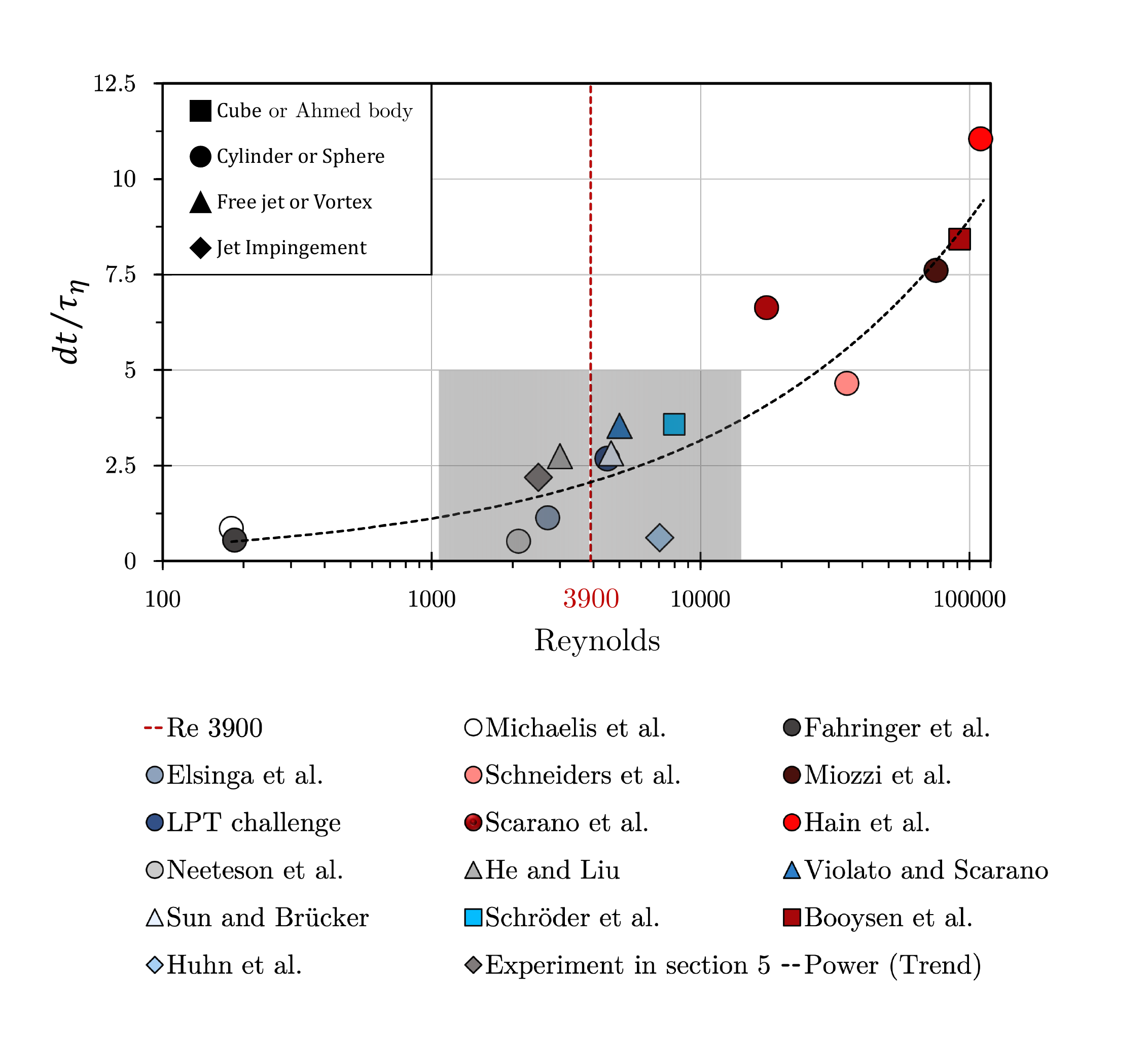}
\caption{\label{fig:fig5} Synthetic time scale selection with respect to recent similar wake flow experimental studies and different Reynolds numbers. Each symbol represents a family of flow configuration \citep{Elsinga2006TomographicVelocimetry,Cfdforpiv.dlr.de20203rdTracking,Neeteson2016Pressure-fieldData,Sun2017InvestigationTomo-PIV,Huhn2018Time-resolvedTracking,Michaelis2006APIV,Schneiders2016Large-scaleTracers,Scarano2015OnExperiments,He2017ProperMeasurements,Schroder2020TheSimulation,Fahringer_2015,Miozzi2015IntegratingCylinder,Hain2008TomographicPlate,Violato2011Three-dimensionalJets,Booysen2019Shake-the-BoxBubbles}
}
\end{center} 
\end{figure}

\citet{Schanz2016Shake-The-Box:Densities} computed the temporal scale as a function of the original experiment time sampling rate resulting in a mean 3D particle displacement of around $6$ pixels for the synthetic data analysis. However, the temporal scale selection requires satisfying the real experiment condition and should be characterised by the flow physics. We defined the temporal resolution as a ratio of turbulence time scales. The non-dimensional form of the temporal scales can be written as $dt/dt_{\rm DNS}$, $dt/\tau_{\eta}$, and $dt/T_D$, 
where $dt$ is the temporal scale for either synthetic or experiment study, and $dt_{\rm DNS}$, $\tau_{\eta}$ and $T_D$ are the DNS, Kolmogorov and integral time scales, respectively. By defining $T_D=D/U_\infty$, we estimate the Kolmogorov time scale from ratio of the largest to smallest time scales as $T_D/\tau_{\eta}\sim Re^{1/2}$. To mimic the real experiment condition, we gathered similar wake flow studies (mainly cylinder wake flows) and plotted the relations of $dt/\tau_{\eta}$ with their Reynolds numbers in Fig.~\ref{fig:fig5}. It is shown that low Reynolds number experiments can resolve the Kolmogorov time scale \citep{Michaelis2006APIV}. The achievable temporal scale increases with the Reynolds number, mainly because the data acquisition frequency limits the experiments with the order of $1-3 \ \text{kHz}$ for the majority of studies mentioned in Fig.~\ref{fig:fig5}. In our synthetic case, the DNS time step is roughly $20$ times smaller than the Kolmogorov time scale. Such a case with every $10$ DNS time step ($dt/\tau_{\eta}=0.47$) can resolve the Kolmogorov time scale, but it is unlikely to achieve such a high acquisition rate in practice, particularly when the fluid is air. Particle trajectories are very smooth with small displacements in this case. However, according to Fig.~\ref{fig:fig5}, the temporal scale $dt/\tau_{\eta}$ stays relatively large, between $2.5$ and $5$ for studies close to a Reynolds number of $3900$ that approximately equals to every $50$ and $100$ DNS time steps. As listed in Table \ref{tab:table2}, we chose three low, moderate, and high time steps starting from every $10$, $50$, and $100$ DNS time steps. It is worth mentioning that it takes $1333$ and $6667$ DNS time steps to reach one integral time scale and one vortex shedding, respectively. 


\subsubsection{Noise ratio}
In this paper, we created noise (i.e., false particles) in the vicinity of true positions. A false particle is randomly distributed around the true position with a maximum displacement radius. Noise ratio (NR) of $0.1$ means $10 \%$ of true particles at every time step have false particles in their vicinities. A false particle around a single track also impacts the track detection accuracy for other neighbour tracks, particularly in dense and intersection situations. In this study, we created three noise ratios, $0$, $0.1$, and $0.2$. 

\begin{table}
\caption{\label{tab:table1} Synthetic particle concentration selection in terms of particles per cubic Kolmogorov and integral scales.}
\begin{ruledtabular}
\begin{tabular}{l c c c}

Case & particles & $\text{pp}\eta^3$  & $\text{pp}D^3$    
\rule{0pt}{10pt} \\

\noalign{\smallskip}\hline\noalign{\smallskip}
 
Low & $2,000$ & $ 10^{-5}$ & 139  \\ 
Moderate & $20,000$ & $ 10^{-4}$  & 1386  \\ 
High & $200,000$ & $ 10^{-3}$  & 13861  \\

\end{tabular}
\end{ruledtabular}
\end{table}



 





 



\begin{table}
\caption{\label{tab:table2} Synthetic time step selection in terms of DNS, Kolmogorov, and integral time scales.}
\begin{ruledtabular}
 \begin{tabular}{l c c c}

 Case & ${dt}/{{dt}_{\rm DNS}}$ & ${dt}/{\tau_{\eta}}$  & ${dt}/{T_D}$     
\rule{0pt}{10pt} \\

\noalign{\smallskip}\hline\noalign{\smallskip}
 
Low & $10$ & $0.47$ & $0.01$  \\ 
Moderate & $50$ & $2.34$  & $0.04$  \\ 
High & 1$00$ & $4.68$  & $0.08$  \\ 

\end{tabular}
\end{ruledtabular}
\end{table}

\subsection{Evaluation and sensitivity analyses}
\label{sec:4_2}


In this section, the performance of the LCTI is assessed and compared against ETI and 4BE-NNI methods. Quantitative accuracy and sensitivity analyses are performed for different particle concentrations, temporal scale, and noise ratio compared to the ground truth trajectories (see Section \ref{sec:4_1}). After the initialisation step, the trajectory of each particle is classified as either untracked, wrong, or true track. A dominant number of true tracks would ease the 4D-PTV process to converge within short time steps. It is worth mentioning that a 4D-PTV process is converged if all particles inside the domain are tracked.



\begin{figure*}
  \includegraphics[width=1\textwidth]{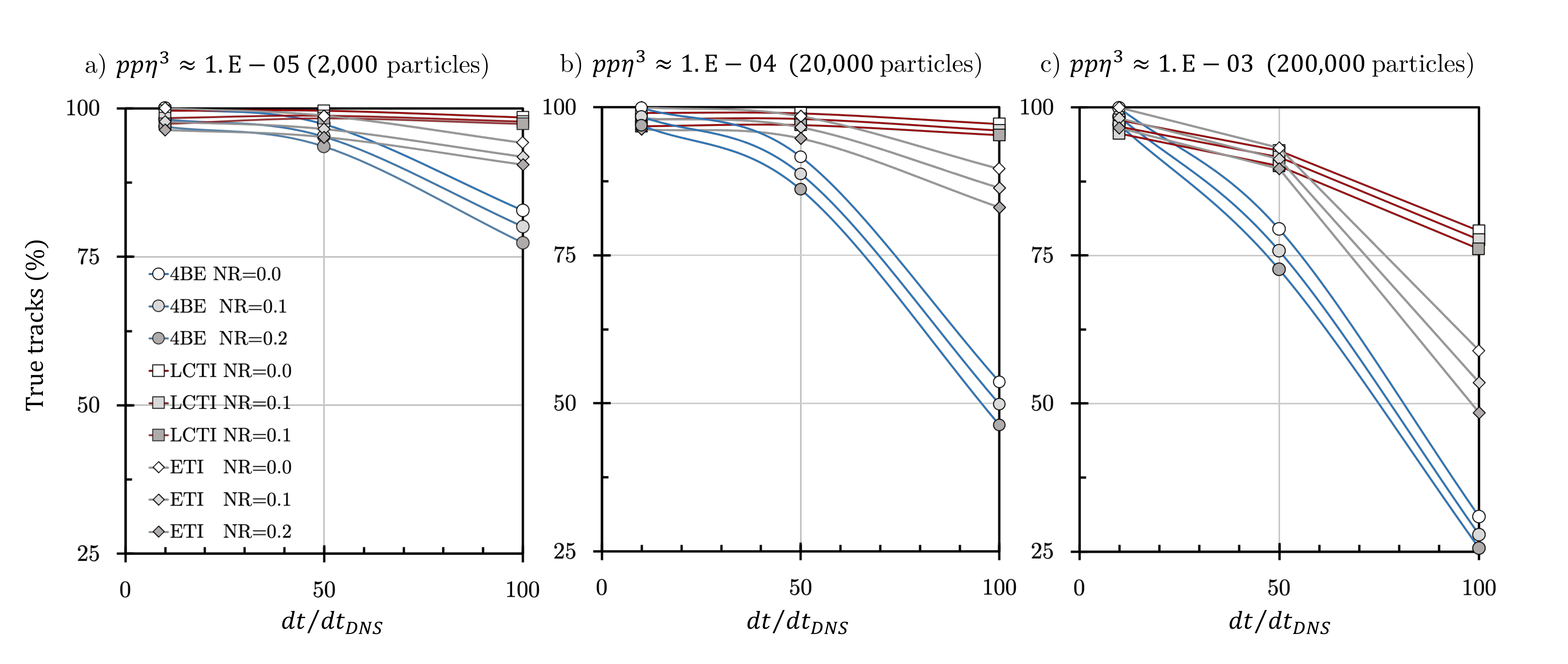}
\caption{\label{fig:fig6}  Comparison performances of three track initialisation techniques in terms of fraction of true particle detection by changing each characteristic parameter. Particle concentration is increasing from a) to c).}
\end{figure*}

Fig.~\ref{fig:fig6} compares the ratio of detected true tracks obtained from LCTI, ETI, and 4BE-NNI techniques with different characteristic parameters of $\text{pp}\eta^3$, $dt/\tau_{\eta}$, and NR. We assessed each technique based on $3\times3\times3$ scenarios representing low, moderate and high levels of each characteristic parameter (see Table \ref{tab:table1} and Table \ref{tab:table2}). All three techniques performed equally well for low particle concentration and small temporal scale by reconstructing over $95\%$ of true tracks (see Fig.~\ref{fig:fig6}a). Tracking is not difficult under such scenarios characterised by small particle displacement with relatively large spatial distance between neighbour tracks. By increasing $dt/\tau_{\eta}$ in low particle concentration cases, the ratio of detected true tracks drops down to approximately $90 \%$ and $75 \%$ in ETI and 4BE-NNI techniques respectively, whereas LCTI remains stable. The large relative distance between neighbour trajectories due to low particle concentration reduces the ambiguity in finding possible tracks. Results of low particle concentration cases for all three techniques show that the 4D-PTV process has to recover less than $25 \%$ of remaining untracked particles to converge in the worst case; thereafter, a short convergence time is expected. The ratio of true tracks drops linearly by increasing the noise ratio (NR) for all techniques with approximately the same order of magnitude. We found that the drop in the detected true tracks caused by the noise ratio (NR) is nearly independent of the other two characteristic parameters. Comparing the three characteristic parameters reveals that the temporal scale has the most deterministic impact on the detected true tracks for all techniques. If $dt/\tau_{\eta}$ stays low, all techniques can cover over $95 \%$ of true tracks regardless of the noise ratio and the particle concentration. Low temporal scale value means that particle displacements are very small. Even two frame nearest neighbour techniques can reconstruct the majority of true tracks. However, in more realistic conditions, when the temporal scale $dt/\tau_{\eta}$ is high, LCTI performed significantly better than the other two techniques. Particle concentration also plays an essential role in the initialisation performance. The ratio of detected true tracks for LCTI stays consistently above $95 \%$ in particle concentrations of $10^{-4} \text{pp}\eta^3$ (see Fig.~\ref{fig:fig6}b). In a worse scenario, when the particle concentration is $10^{-3} \text{pp}\eta^3$ associated with high temporal scale, LCTI still can recover over $75 \%$ of true tracks that is considerably higher than ETI and 4BE-NNI with roughly $50 \%$ and $25 \%$ true tracks, respectively (see Fig.~\ref{fig:fig6}c).  

\begin{figure*}

  \includegraphics[width=1\textwidth]{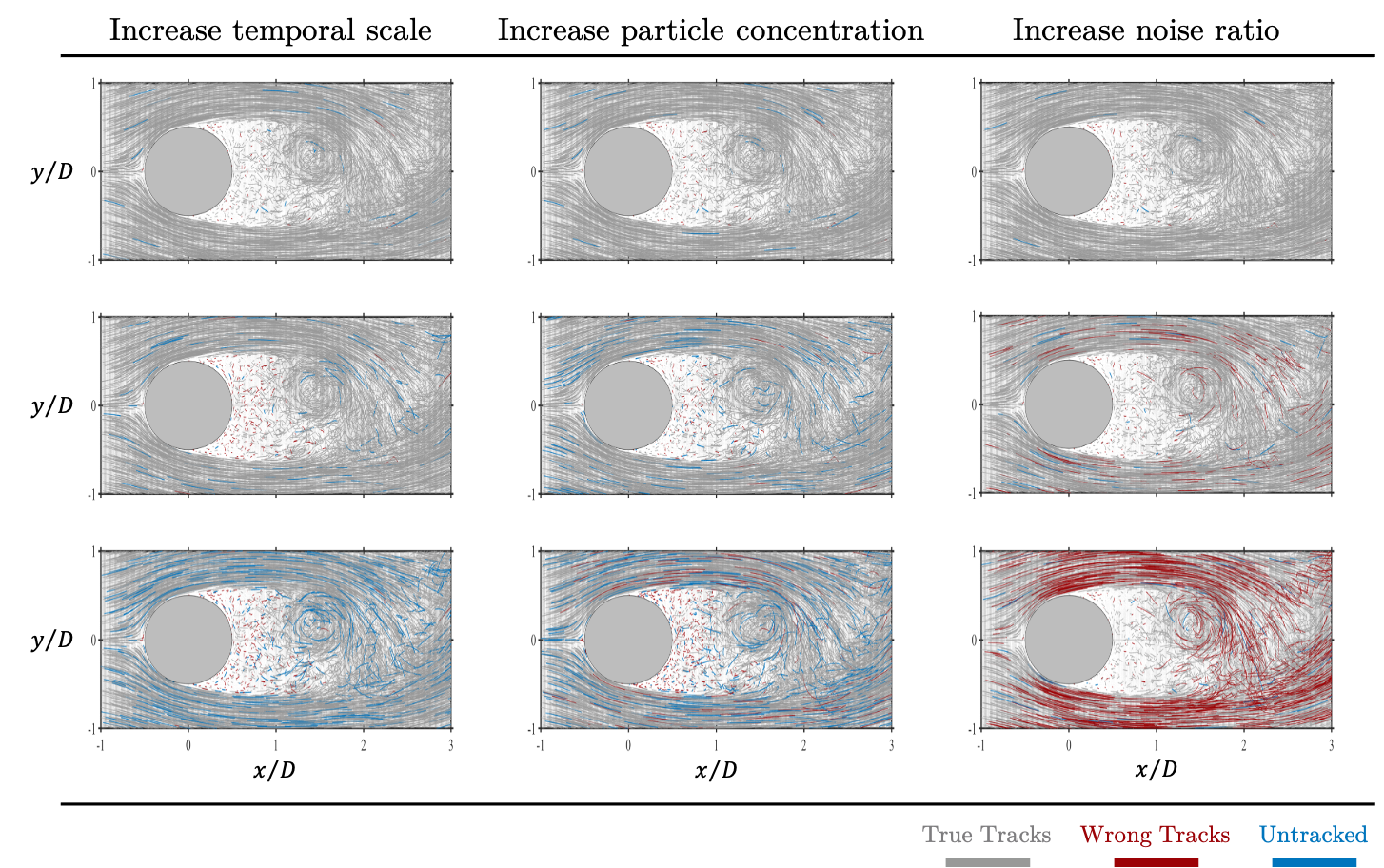}
\caption{\label{fig:fig7}  Sensitivity analysis of LCTI compared with DNS ground truth by increasing noise level, particle concentrations, and temporal scale from top to bottom. Tracks in blue and red colour represent untracked and wrong trajectories respectively. Gray tracks are true trajectories that the algorithm built (numbers of grey tracks are scaled down for having a clear view of red and blue tracks).}
\end{figure*}

\begin{figure}
\centering
  \includegraphics[width=0.5\textwidth]{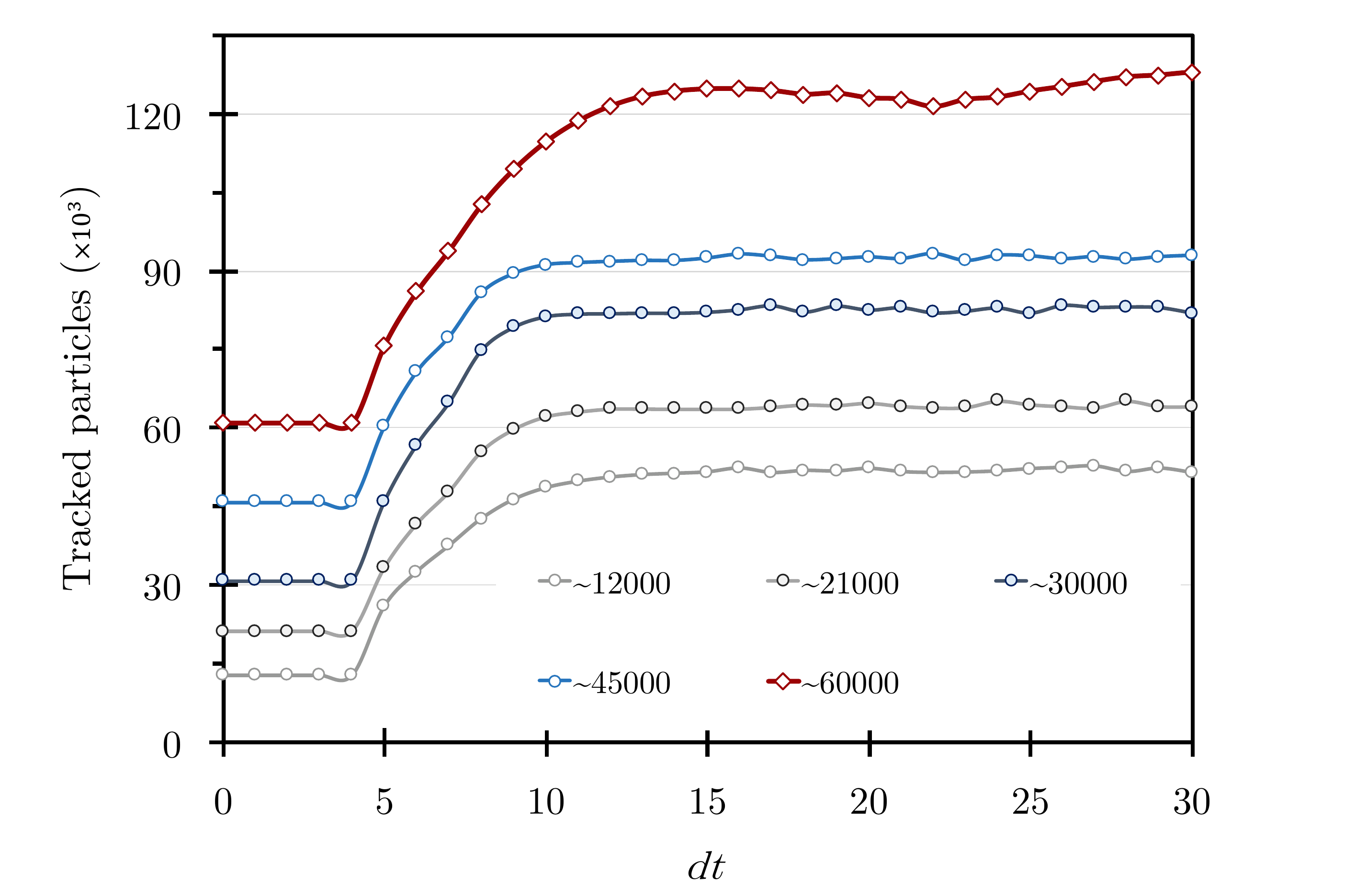}
\caption{\label{fig:fig8} Sensitivity of the 4D-PTV convergence to the number of initialised tracks for the LPT challenge high-density case at $0.12\ \text{ppp}$ over first $30$ time steps. Number of initialised tracks after the first four frames varies from $\sim 12,000$ to $\sim 60,000$}
\end{figure}
The synthetic data evaluation showed that LCTI systematically outperforms other competing techniques. However, it should be mentioned that the cost of this achievement is expensive because it computes the coherency for every possible tracklet. We found that LCTI requires roughly $4$ times more computation time on a single CPU core than a classic four frame based initialisation technique without any post-treatment. To this end, an appropriate initialisation technique should be chosen depending on the measurement condition. As an example, there is no need to perform a sophisticated initialisation technique if the temporal scale (i.e., sampling rate) and the particle concentration are low in comparison with the turbulence scales. It is important to note that using sophisticated initialisation techniques such as LCTI is crucial to prevent 4D-PTV failure in challenging cases with a large temporal scale and high particle concentration. 
\begin{figure*}
\begin{center} 
  \includegraphics[width=0.85\textwidth]{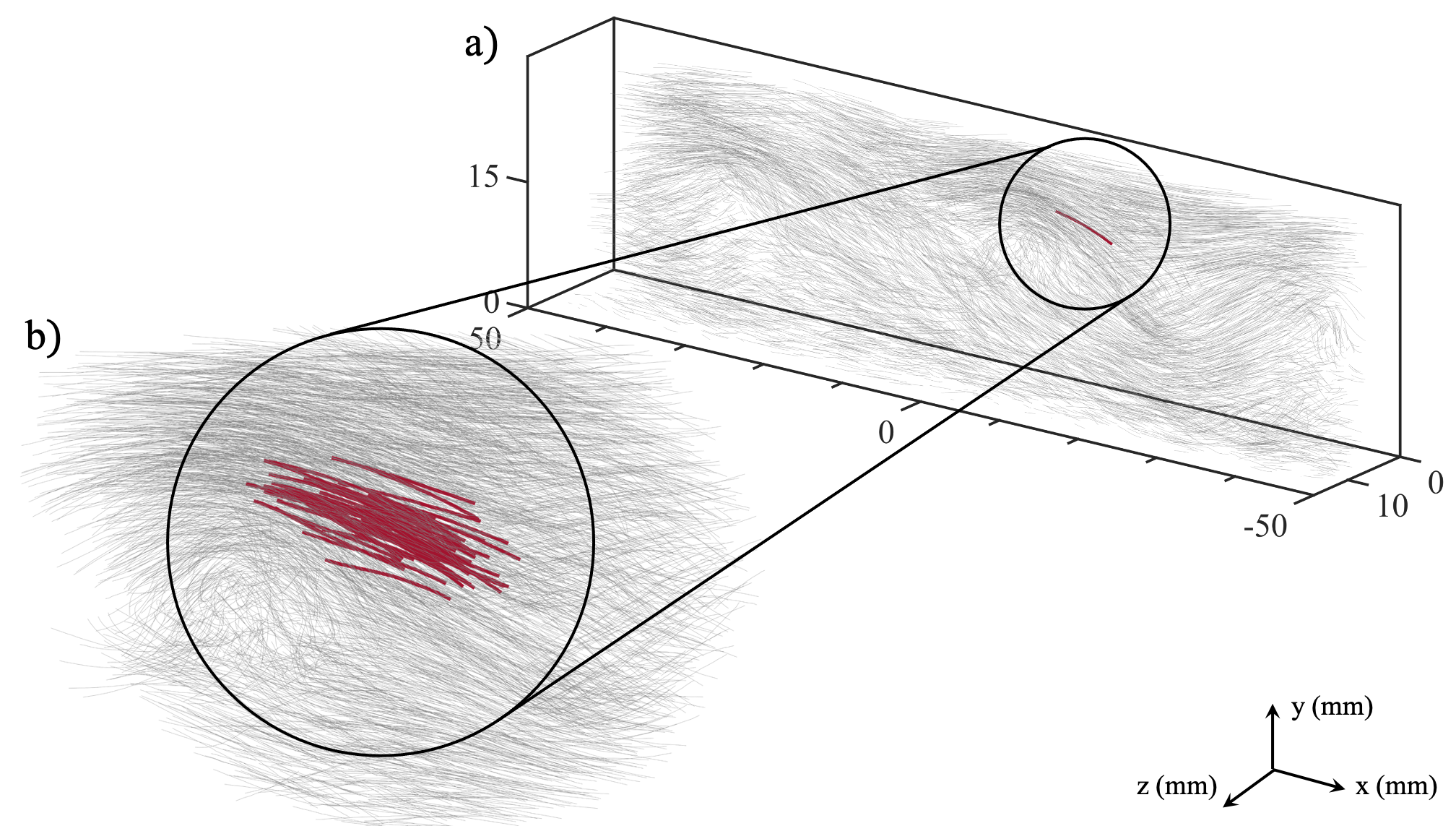}
\caption{\label{fig:fig9} LCTI trajectory results for the LPT challenge wake flow at $0.12\ \text{ppp}$ (\citep{Cfdforpiv.dlr.de20203rdTracking,RahimiKhojasteh2020LagrangianInitialisation}); a) slice view  in $y$ direction of particle trajectories in grey and the target track in red b) Cluster of coherent tracks with the target track in red }
\end{center}
\end{figure*}
Apart from the initialisation performance based on the number of true tracks, we performed further parametric analyses on LCTI to determine how untracked and wrong trajectories are sensitive to the characteristic parameters. We started from a base case with $10^{-4} \ \text{pp}\eta^3$ particle concentration, $dt/\tau_{\eta}=2.34$, and zero noise ratio. Afterwards, we increased each parameter separately until the fraction of true tracks drops down to $80 \%$. The remained $20 \%$ is a mix of untracked and wrong trajectories. Lagrangian flow maps for increasing of each characteristic parameter are shown in Fig.~\ref{fig:fig7}. If we increase the temporal scale, both untracked and wrong tracks increase as the number of true tracks drops. Under such a scenario, the particle displacements are large and comparable with local distances between the neighbour particles. Therefore, less information between two-time steps is available, which drastically increases the possibility of having untracked particles. Depending on the region of the flow, the majority of untracked trajectories (blue colour in Fig.~\ref{fig:fig7}) exist around the two high shear sides of the wake region. 
Although the number of untracked trajectories dominate the whole Lagrangian flow map, more wrong tracks (red colour in Fig.~\ref{fig:fig7}) than untracked trajectories are observed inside the wake region. By increasing the particle concentration, both untracked and wrong trajectories raise homogeneously with nearly the same proportion through the Lagrangian flow map. Interestingly, the number of wrong tracks is still comparably larger than the number of untracked particles inside the wake region. It shows that regional flow behaviours can have a direct impact on the initialisation performance. The impact of increasing the noise ratio is also shown in Fig.~\ref{fig:fig7}. We found a dominated number of wrong tracks in all regions with increased noise ratio. In this study, we created the noise (i.e., false particles) in the vicinity of true particles that causes more initialisation ambiguities. Due to this reason, LCTI yields more wrong trajectories with increased noise ratio. Besides, wrong indexing of a track takes away at least one true or false particle that may propagate the wrong detection to another neighbour track too. In practice, for 4D-PTV,  wrong initialised tracks increase the failure risk and need to be appropriately eliminated in a subsequent prediction-optimisation step. The untracked particles, on the other hand, require more iteration and convergence time.

Sensitivity analysis in Fig.~\ref{fig:fig7} implies that LCTI is likely to have more wrong tracks inside the wake region with increased noise ratio, particle concentration, and temporal scale. However, different behaviours have been seen depending on the characteristic parameters in overall. The high turbulence intensity level at Reynolds $3900$ creates small flow structures inside the wake region. Consequently, particles inside this region are coherent with a small number of neighbours and quickly change to different coherent clusters. This behaviour brings more complexity for the coherency detection that might be the reason for having a dominating number of wrong tracks, despite relatively small particle displacements. \citet{Schanz2020Shake-The-BoxVT-STB} observed a similar issue when particles of a specific region move slowly compared to the rest of the domain. One approach for improving trajectory in such regions is performing temporal filtering schemes by adjusting the temporal scale to track the slowest particles \citep{Schanz2020Shake-The-BoxVT-STB}.




\subsection{LPT challenge}
\label{sec:4_3}



LCTI was implemented into KLPT \citep{yang:hal-03212696} to run the whole 4D-PTV process (see Fig.~\ref{fig:fig1}). KLPT featured by LCTI (KLPT-LCTI) was examined on the time-resolved data from the LPT challenge \citep{Cfdforpiv.dlr.de20203rdTracking,RahimiKhojasteh2020LagrangianInitialisation} at four particle densities (i.e., concentrations) from $0.005 \ \text{ppp}$ up to $0.08 \ \text{ppp}$. The challenge cases were obtained from the wall-bounded wake flow behind a cylinder at a momentum thickness Reynolds number $Re_\theta$ of around $4500$. In terms of turbulence length scales, particle concentration of the mentioned four cases varied between $2 \times 10^{-7} \ \text{pp}\eta^3$ and $3 \times 10^{-6} \ \text{pp}\eta^3$. The domain of interest was set at  $100 \ \text{mm} \times\ 50 \ \text{mm} \times\ 30 \ \text{mm}$ downstream of the cylinder. The image acquisition rate was equal to $600 \ \mu s$, resulting in $dt/\tau_{\eta}=2.68$ temporal scale. At the lowest $\text{ppp}$, the proposed method managed to reconstruct over $99 \%$ of true particles accurately. Percentage of true particles maintained over $99 \%$ for higher densities (i.e., $0.025$, $0.05$, and $0.08 \ \text{ppp}$). Accordingly, missed and ghost particles were less than $1 \%$. The case studies of the LPT challenge revealed that the positional Root-Mean-Square error (RMSE) of KLPT-LCTI increased linearly with $\text{ppp}$, but it remained below $0.0041 \ \text{mm}$ for all four particle densities. This illustrates the reliable performance of the LCTI at particle densities lower than $0.08 \ \text{ppp}$, knowing that most of the 4D-PTV real experiments perform at $0.05 \ \text{ppp}$ particle density or lower.


For densities higher than $0.08 \ \text{ppp}$, a more accurate initialisation technique could prevent the 4D-PTV algorithm from failing or improve its convergence speed. We highlighted that KLPT featuring LCTI succeeded in reconstructing tracks at the density of $0.12 \ \text{ppp}$ while KLPT featuring NNI failed to converge. Questions have been raised about the 4D-PTV sensitivity to the number of initialised particles at the beginning. We illustrated this issue in the LPT challenge case with $0.12 \ \text{ppp}$ and over $120,000$ particles. As shown in Fig.~\ref{fig:fig8}, the KLPT-LCTI process reaches no more than $85,000$ (i.e., $70 \%$) final tracks if the process starts with any number below $30,000$ initialised tracks. However, starting with $60,000$ initialised tracks leads to cover over $99 \%$ of final trajectories after $30$ time steps at $0.12 \ \text{ppp}$. The evidence from this study indicates that the number of initialised tracks is one deterministic contributor to the 4D-PTV convergence at high-density scenarios. Without a proper track initialisation algorithm, a 4D-PTV scheme would not be able to eventually recover the majority of tracks.


\section{Experiment study}
\label{sec:5}
To demonstrate the potential of the LCTI in a practical configuration, we used the data from a volumetric experiment of liquid-liquid jet impingement on a circular wall carried out in our laboratory at a Reynolds number equal to $2500$. Perpendicular impinging the jet into the wall creates significant deceleration inside the jet core. The deceleration and directional 3D motions of particles are featured by multi-vortex rings around the jet and secondary vortex structures after the impingement. 

\begin{figure}
\begin{center} 

  \includegraphics[width=0.5\textwidth]{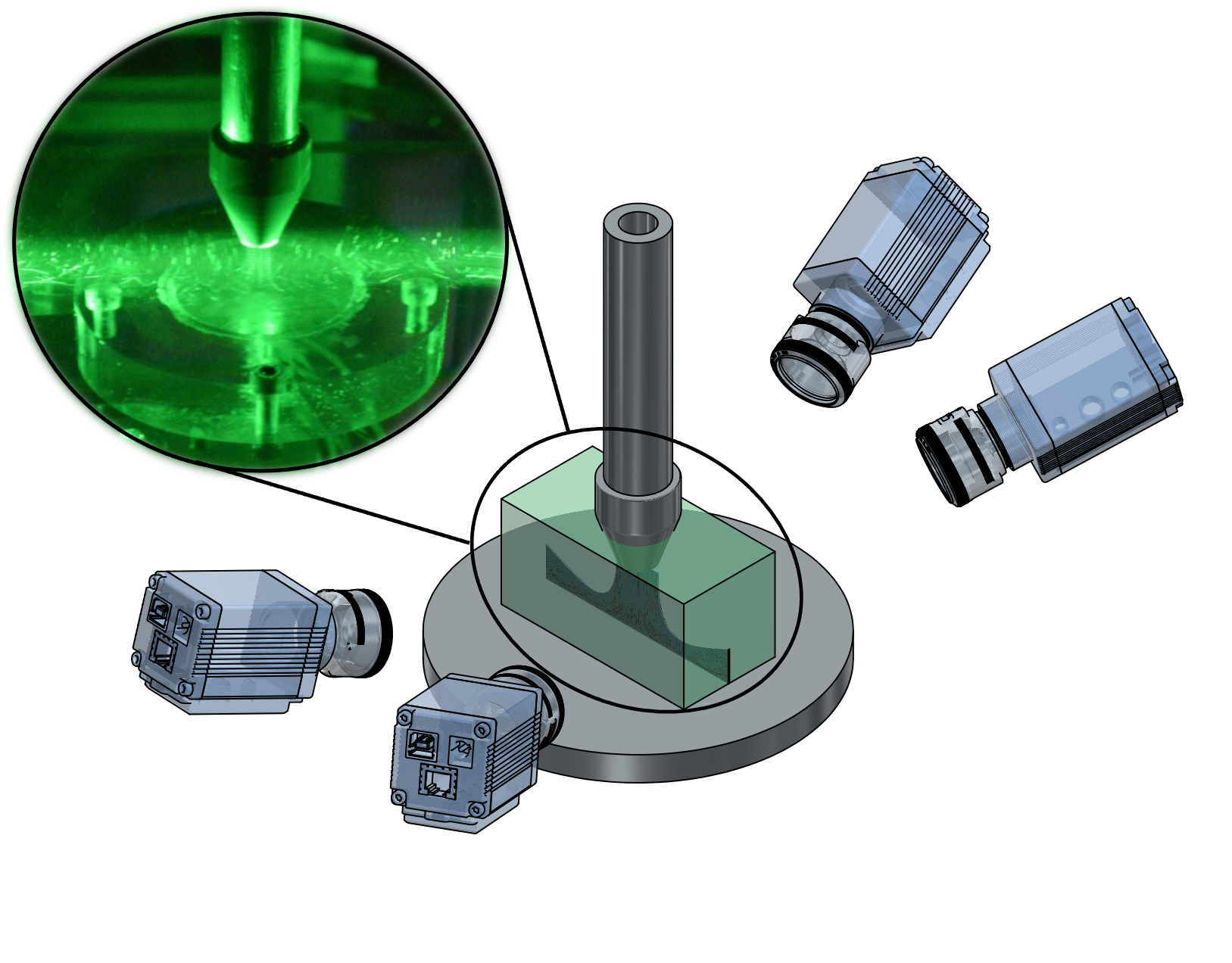}


\caption{\label{fig:fig10} 3D schematic view of the jet impingement experimental, with snapshot of the experiment in zoom balloon.}
\end{center}
\end{figure}

Fig.~\ref{fig:fig10} shows the schematic view of the experiment setup. Four Phantom $M310$ cameras with $1200\times800$ resolution and the maximum frequency of $3 \ \text{kHz}$ empowered. Nikon $105 \ \text{mm}$ macro $\text{F} 2.8$ (aperture was set to $\text{F} 22$) lenses were employed. As shown in Fig.~\ref{fig:fig10}, two cameras were positioned in $24$ degree with forward scatter light, and the other two cameras were in backward scattering at $13$ degree. $15 \ \text{mJ}$ LDY $300$ laser was set for the illumination with $1  \ \text{kHz}$ frequency ($dt/\tau_{\eta}=2.18$) and converted into the rectangular light volume. The measurement volume was $16 \ \text{mm} \times 51 \ \text{mm} \times 35 \ \text{mm}$ ($29 \ \text{cm}^3$).  The seeding particles were hollow glass spheres $9-13\ \mu m$ in diameter, and $1.1 \ \text{g}/\text{cm}^3$ density, and the particle concentration was approximately $0.03 \ \text{ppp}$ that was equivalent to $2 \times 10^{-6} \ \text{pp}\eta^3$. More details in this experiment can be found in \citet{yang:hal-03212696}.

\begin{figure}
  \centering
  \subfloat[]{\includegraphics[width=0.5\textwidth]{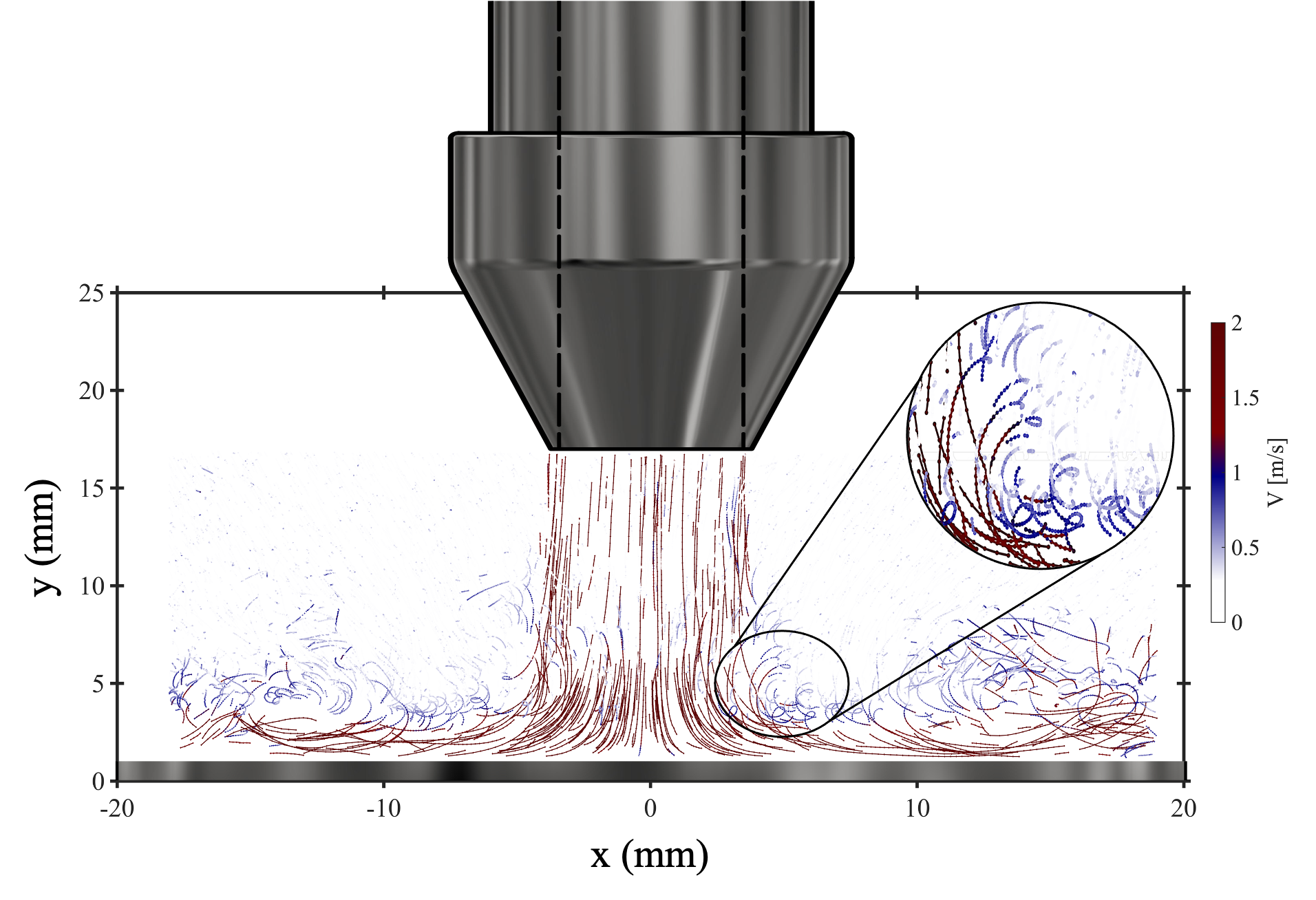}\label{fig:f1}}
  \hfill
  \subfloat[]{\includegraphics[width=0.5\textwidth]{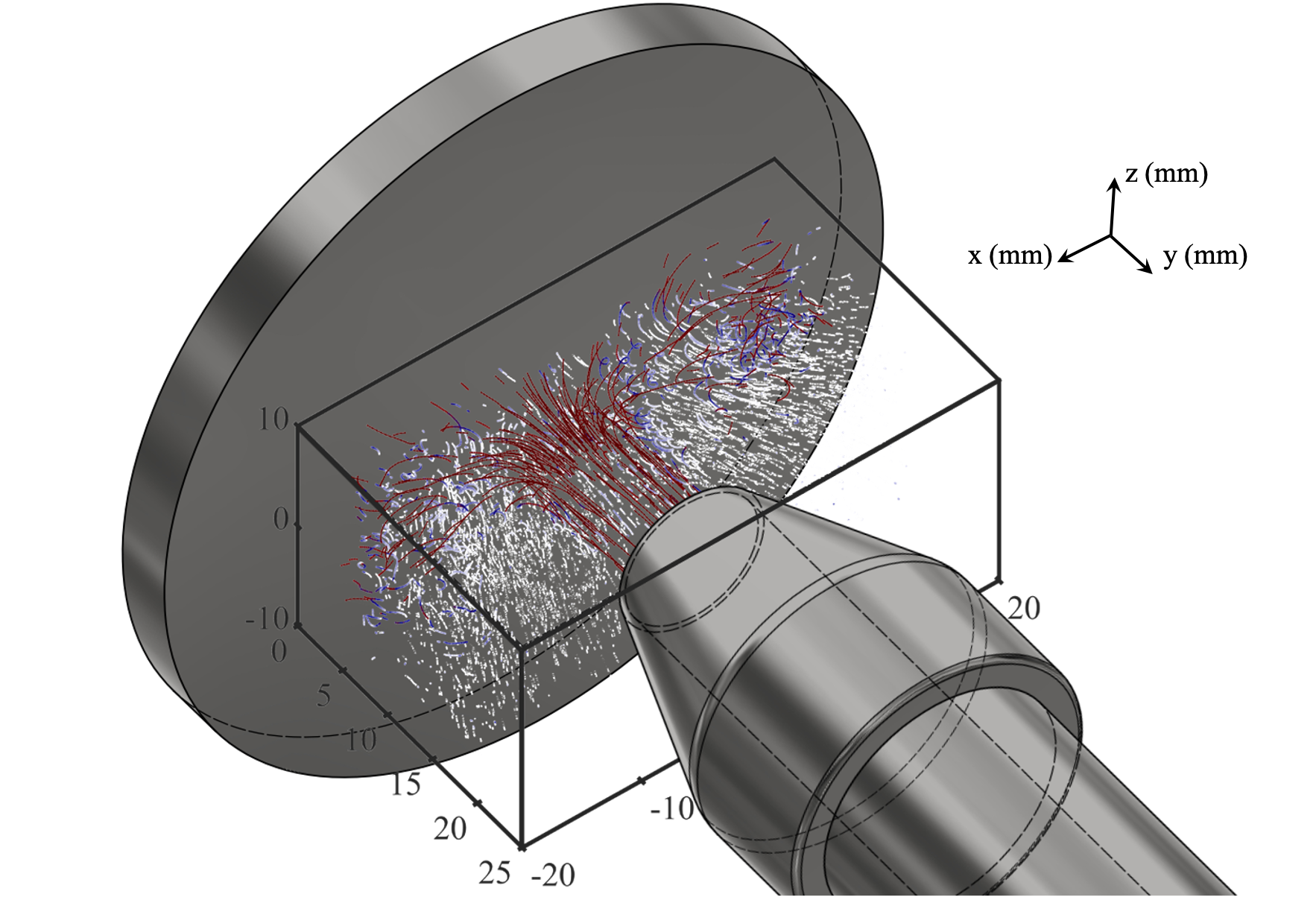}\label{fig:f2}}

  \caption{\label{fig:fig11} a) Side view of the trajectories coloured by their velocity magnitudes at $0.03 \ \text{ppp}$. b) 3D view of the trajectories in the Jet impingement. Low-velocity tracks away from the jet core are filtered for clear qualitative view thanks to the colorbar}

\end{figure}

In this section, we performed the 4D-PTV process without prediction and optimisation parts to particularly demonstrate the LCTI performance. Therefore, the particle positions reconstructed by IPR were employed in $20$ consecutive time steps and followed by multi four-frame LCTI processes. Trajectory results obtained from LCTI are shown in Fig.~\ref{fig:fig11}. Away from the jet impingement region, the trajectories are relatively smooth with small displacements. As explained earlier in \ref{sec:4_1}, there is no critical issue in such regions, and a simple initialisation technique can track the majority of particles. Small trajectories away from the jet, therefore, are omitted in Fig.~\ref{fig:fig11} to concentrate on the most complex regions. As the jet impinges on the wall, trajectories decelerate highly, turn alongside the wall, but still tend to keep their coherent local motions. The proposed method effectively detected these coherent trajectories, although complex behaviours exist. We found that the trajectories around the jet tend to circulate (see the zoom balloon in Fig.~\ref{fig:fig11}). This reveals strong evidence of the particle coherent motions impacted by the vortex rings. Near wall trajectory reconstructions show signatures of the secondary vortices where particles bounce back. The results of this experiment test case support the idea that each particle can temporally and spatially behave in coherence with a group of other neighbour particles. The second significant finding was that the proposed initialisation technique could cover most tracks in complex and high gradient regions associated with 3D directional dynamics.

\section{Conclusions and outlook}
\label{sec:6}

We proposed a novel technique to reconstruct tracklets from four (or multi) frames by leveraging temporal and local spatial coherency of neighbour tracks. These tracks should be consistent with the neighbour coherent motions bounded by LCS ridges. To assess the LCTI performance in various conditions, we have created an open-access synthetic dataset for the wake flow downstream of a smooth cylinder obtained from DNS at a Reynolds number equal to $3900$. Future studies by interested readers should focus on using the current open access Eulerian velocity and pressure data as well as synthetic Lagrangian trajectories for algorithm evaluation purposes. We transported particles using the trilinear spatial interpolation scheme followed by the 4th order Runge Kutta temporal integration. In the current dataset, temporal and spatial scales are reported based on turbulence length and time scales. 


As a result of the parametric study in section \ref{sec:4_2}, we found that a 4D-PTV process requires more sophisticated initialisation techniques if either one of the following conditions meets: 1), Particle displacements between two time steps are relatively large due to either sparse temporal acquisition frequency or high dynamic gradients of particles (velocity and accelerations); 2), The particle concentration is high enough to have a length between neighbour particles with the same order of their trajectory displacement lengths; 3), The reconstructed particle field is noisy. Under these conditions, LCTI tends to detect more true tracks than recent ETI and 4BE-NNI techniques. We found that the temporal scale severely impacts the true track detection yielded by any initialisation techniques. We also analysed the proportions of untracked and wrong trajectories in LCTI. The main results indicate that the noise ratio creates more wrong tracks, the particle concentration homogeneously increases both untracked and wrong tracks, and finally, the temporal scale causes more untracked trajectories. Regardless of the characteristic parameters, we found more wrong trajectories than untracked particles inside the wake region. 
We also applied LCTI to the time-resolved dataset in the LPT challenge after integrating LCTI to a 4D-PTV scheme, KLPT. Our KLPT-LCTI scheme has achieved state-of-the-art performance for densities up to $0.08 \ \text{ppp}$ \citep{Cfdforpiv.dlr.de20203rdTracking,RahimiKhojasteh2020LagrangianInitialisation}. LCTI is helpful for high particle density data where the portion of initialised tracks after the initialisation stage directly impacts the 4D-PTV's convergence performance. At $\text{ppp}=0.12$, KLPT featuring a simple nearest neighbour initialisation scheme fails to yield any valid results. LCTI, on the contrary, can recover much more tracks from the starting four frames, therefore provides a more accurate initialised track field. Besides, LCTI also contributes to following every frame by bringing more new tracklets (length less than 4) from triangulated particles on residual images into the tracked poll (see Fig.~\ref{fig:fig1}). This finding is in agreement with the synthetic analysis part in section \ref{sec:4_2} showing the importance of having an advanced initialisation for dense (i.e., high particle concentration) conditions. 
LCTI was then tested on the jet impingement experiment. Although the flow was characterised by complexities such as 3D directional motion and trajectory intersections, LCTI successfully reconstructed the majority of tracks solely on particle fields reconstructed by IPR, without the necessity of further prediction and optimisation processes. The trajectory results also showed particles coherent motions in large scale flow motions such as vortex rings, impinging area, and secondary vortices. This comprehensive study has demonstrated that coherency based track initialisation is a robust approach to reconstruct tracks even in complex situations. The proposed technique can be used either as an embedded module for the 4D-PTV process, or as a standalone four-frame-based tracker.
LCTI showed that additional physics-based information increases the accuracy and robustness of the initialisation part. For further investigations, the FTLE function can be replaced with advanced LCS detection algorithms 
(see, e.g., \citet{Filippi2020AnFlows}) or other coherent motion detection techniques such as Coherent Structure Colouring \citep[CSC]{Martins2021DetectionColouring}. More research is needed to understand and track clusters of coherent particles in real experiments.

\begin{acknowledgements}
\label{sec:7}
The authors would like to greatly thank the LPT challenge organising committee, particularly Dr Andrea Sciacchitano from TuDelft university, for providing the time-resolved challenge results. A special thanks from the authors is given to Anthony Guibert (a former assistant engineer at INRAE) who conducted the impingement experiment in $2016$. The authors would finally like to thank EPSRC in the UK for the computational time made available on the UK supercomputing facility ARCHER2 via the UK Turbulence Consortium (EP/R029326/1).
\end{acknowledgements}
\nocite{*}
\bibliography{LCTI}
\end{document}